\documentclass{aa}
\usepackage{graphicx}
\usepackage{psfig}
\usepackage{rotating}
\usepackage{amssymb}
\usepackage{latexsym}
\usepackage{psfig}
\usepackage{epsfig}
\usepackage{graphics}

\begin{document}

\title{	Galaxy morphology and evolution from SWAN \\ Adaptive Optics imaging
      \thanks{Based on observations collected at the European Southern
        Observatory, Chile under programs 70.B-0649, 71.A-0482 and 073.A-0603.} 
}
\author{}
\author{G. Cresci \inst{1} \and R.~I. Davies \inst{2} 
  \and A.~J. Baker \inst{3,4} \and F. Mannucci \inst{5} 
  \and M.~D. Lehnert \inst{2} \and T. Totani \inst{6} 
  \and Y. Minowa \inst{7} 
}
\institute{INAF - Osservatorio Astrofisico di Arcetri,
    	Largo E. Fermi 5, I-50125, Firenze, Italy
        \and Max-Planck-Institut f\"ur extraterrestrische Physik,
        Postfach 1312, D-85741 Garching, Germany        
        \and Jansky Fellow, National Radio Astronomy Observatory
        \and Department of Astronomy, University of Maryland, College 
        Park, MD 20742-2421, United States
	\and CNR-Istituto di Radioastronomia, Largo E. Fermi 5, I-50125, 
	Firenze, Italy
	\and Department of Astronomy, Kyoto University, Kitashirakawa, 
	Kyoto 606-8502, Japan
	\and Institute of Astronomy, School of Science, University of Tokyo, 
	2-21-1 Osawa, Mitaka, Tokyo 181-0015, Japan
        }

\offprints{G. Cresci \\
    \email{gcresci@arcetri.astro.it}}

\date{Received / Accepted}

\abstract{We present the results from adaptive optics (AO) assisted
imaging in the $K_\mathrm{s}$ band of an area of $15\ \textrm{arcmin}^2$ for 
SWAN (Survey of a Wide Area with NACO).  We derive the high resolution 
near-IR morphology of $\sim 400$ galaxies up to $K_\mathrm{s} \sim 23.5$ in 
the first 21 SWAN fields around bright guide stars, carefully taking into 
account the survey selection effects and using an accurate treatment of the 
anisoplanatic AO PSF.  The detected galaxies are sorted into two morphological 
classes according to their S\'ersic index.  The extracted morphological 
properties and number counts of the galaxies are compared with the predictions 
of different galaxy formation and evolution models, both for the whole galaxy
population and separately for late-type and early-type galaxies. This is one 
of the first times such a comparison has been done in the near-IR, as AO observations 
and accurate PSF modeling are needed to obtain reliable morphological 
classification of faint field galaxies at these wavelengths.  
For early-type galaxies we find that a pure luminosity evolution model, 
without evidence for relevant number and size evolution, better reproduces the observed 
properties of our $K_\mathrm{s}$-selected sample than current semi-analytic models 
based on the 
hierarchical picture of galaxy formation. In particular, we find that the observed 
flattening of elliptical galaxy counts at $K_\mathrm{s}\sim20$ is quantitatively in 
good agreement with the prediction of the pure luminosity evolution model that was 
calculated prior to the observation.
For late-type galaxies, while both models are able to reproduce the number counts, 
we find some hints of a possible size growth. 
These results demonstrate the unique power of AO 
observations to derive high resolution details of faint galaxies' morphology 
in the near-IR and drive studies of galaxy evolution.
       }


\maketitle


\section{Introduction}

One of the main objectives of modern astrophysics is understanding the process 
of galaxy formation and evolution. The best way to tackle this issue is 
studying the properties of galaxies observed at the epoch of their formation 
and early evolution, such as their stellar population, history of mass assembly, 
morphology, metallicity and interplay with the intergalactic medium. 
However, disentangling these processes in nearby systems 
is already extremely difficult, and the challenge is even greater at higher 
redshift, where sources are compact in size ($\sim 0.1\arcsec - 0.3\arcsec$) 
and larger galaxies are rare (e.g., Bouwens et 
al. \cite{bouwens04}). To resolve and study the details of high-redshift 
galaxies using ground based telescopes, which can provide larger samples and 
deeper observations than space-based observations, it is
necessary to overcome the blurring effects of the atmosphere through
the use of adaptive optics (AO) systems. These can allow ground-based 
telescopes to operate at or near the diffraction limit in the near-infrared 
($\sim 0.07\arcsec$ in $K$ band for an 8\,m telescope), resulting in a high 
angular resolution and a low background in each pixel.

Besides the technical advantages afforded by AO, near-infrared surveys provide 
one of the best opportunities to investigate the cosmic evolution of galaxies 
and their mass assembly.  In particular, $K$-band (2.2\,${\rm \mu m}$) 
selected samples are ideally suited for addressing the problems of galaxy 
formation and evolution.  First, since the rest frame near-IR luminosity is a 
good tracer of the galaxy stellar mass (e.g., Brinchmann \& Ellis 
\cite{brinchmann}; Bell \& de Jong \cite{bell01}; Mannucci et al. 
\cite{mannucci05}), $K$-band surveys allow us to select galaxies according 
to their mass up to $z \sim 1.5$ ($\lambda_{\rm rest} \sim 0.9-1.0\,{\rm \mu 
m}$), rather than suffer strong biases towards star-forming and peculiar 
galaxies like optical surveys (e.g. Drory et al. \cite{drory04}, Fontana 
et al. \cite{fontana04}).  Another strong argument for selecting galaxies 
in the near infrared is that, due to the similarity of the spectral shapes of 
different galaxy types and stellar population ages in the rest frame near-IR 
over a wide redshift range (e.g., Mannucci et al. \cite{mannucci01}), the 
selection of galaxies in the $K$ band is not affected by strong 
$k$-correction effects (e.g., Cowie et al. \cite{cowie94}).  In contrast, 
selection in the $I$ band becomes very type sensitive beyond $z=1$, and the 
situation is even more extreme in the $B$ band, where the fading of early-type 
galaxies is substantial even at modest redshifts.  Thus, near-IR samples do 
not depend as strongly on galaxy type as optically selected ones, which are 
more sensitive to recent and ongoing star formation activity (as they 
sample the rest-frame UV light) and are biased against old and passive or weakly 
star-forming galaxies.

Finally, near-IR surveys are less affected by dust extinction than optical 
ones, making it possible to select highly extinguished star-forming galaxies.  
The observation of the obscured dusty star formation rate is crucial for 
measuring the global star formation history.  Calculations based on the 
observed rest frame UV flux (e.g., Madau et al. \cite{madau96}; Connolly et 
al. \cite{connolly97}) might be significantly underestimated if a large 
fraction of the overall star formation at high redshift takes place in highly 
obscured starburst galaxies (e.g., Steidel et al. \cite{steidel99}; Blain et 
al. \cite{blain02}).

Morphology is one of the most appropiate ways to characterize the properties 
of galaxies, and we will only reach a complete understanding of galaxies by 
deriving the mechanisms responsible for their morphologies. 
In this context, the study of galaxy size, and of the evolution of other galaxy 
properties according to morphological type, have made use mainly 
of the classification derived from deep optical HST imaging (e.g., 
Simard et al. 
\cite{simard99}; Labb\'e et al. \cite{labbe03}; 
Trujillo \& Aguerri \cite{trujillo04};
Pannella et al. \cite{pannella06}), due to the higher angular resolution 
achievable at optical wavelengths with HST.
However, near-infrared morphology is a better tracer of the underlying mass 
distribution, as it is not biased towards recent star formation and 
is less affected by dust obscuration. 
By using adaptive optics, it is now possible to push the analysis of source 
properties (surface density, magnitude, color, morphology, etc.) as a function 
of source size in the near-IR to an entirely new regime, and study sources that are both 
faint \textit{and} compact.  Ample evidence already indicates 
that such source populations do exist -- e.g., a large fraction of the $H_{AB}<21$ sources 
detected by Yan et al. (\cite{yan98}) are still unresolved at the 
$\sim 0.35\arcsec$ resolution provided by HST/NICMOS in the near-IR.  The 
AO-corrected, diffraction-limited, near-IR PSF of an 8\,m telescope is a 
powerful tool to study this kind of object, since the angular resolution
it yields is even higher than can be obtained by HST at this wavelength.

Although the advantages of near-IR AO observations for studying how galaxies 
form and evolve in the early universe are clear, until now there have been 
only a few attempts using natural guide stars (NGS; see 
e.g., Larkin et al. \cite{larkin}; Glassman, Larkin \& Lafreni\'ere 
\cite{glassman}; Steinbring et al. \cite{steinbring04}; Minowa et al. 
\cite{minowa}), due to the very small number of known extragalactic sources 
lying at distances $\Delta \theta \la 30\arcsec$ from bright ($V \la 13$) 
stars needed to correct the wavefront for AO guiding, and to the problems 
arising from the anisoplanaticism of the PSF in AO observations.  The 
prospects for AO cosmology will undoubtedly improve with the widespread 
adoption of laser guide star (LGS) systems, since these impose less stringent 
requirements on the brightness of stars used for tip-tilt correction (e.g., 
Melbourne et al. \cite{melbourne}). However, to overcome the present 
shortage of targets for AO cosmology, it is necessary to identify and 
characterize extragalactic sources in the vicinity of bright guide stars (see 
e.g., Larkin et al. \cite{larkin99}; Davies et al. \cite{davies01}; 
Christopher \& Smail \cite{cresm}). 

We therefore undertook a campaign of seeing-limited near-IR imaging of fields 
selected around stars bright enough for AO guiding ($10.3 \le R \le 12.4$), 
blue ($B-R \le 1.1$, in order to maximize the amount of light on the wavefront 
sensor), lying at high galactic latitude ($|b| \ge 15\deg$, to minimize 
extinction and contamination by foreground stars), and with a declination 
suitable for observations with the ESO Very Large Telescope at low air mass 
($-44\deg \le \delta \le -13\deg$).  A total of 42 southern bright star fields 
(SBSFs) were selected and observed at seeing-limited resolution in 
$K_\mathrm{s}$ band with SOFI at the ESO {\em New Technology Telescope}.  More 
details about the target selection and data can be found in Baker et al. 
(\cite{baker}). The same fields have been followed up at optical wavelengths 
(Davies et al. \cite{davies05}), and are now targets for VIMOS integral field 
optical spectroscopy at the ESO Very Large Telescope (VLT). 

In this paper we present the results of our $K_s$-band AO imaging survey of the first 
21 fields in the framework of SWAN (Survey of a Wide Area with NACO), which is 
the AO-assisted result of these seeing-limited preliminaries. The survey will 
be introduced in the following section, and the observations will be briefly 
described in section \ref{obs}. The data reduction approach will be presented 
in section \ref{reduc}, while the detection criteria and technique will be 
discussed in section \ref{detect}.  The extraction of the morphological 
parameters of the detected galaxies is analyzed in section \ref{morph}, 
and the method used to distinguish between stars and galaxies is described  
in section \ref{stargal}. In 
section \ref{counts} we take into account the selection effects 
present in our data, discuss the completeness of the survey, and show 
the corrected number counts. The number counts and size-magnitude relation of the 
full sample of galaxies and for late and early-type systems separately are 
compared with the predictions of two different galaxy evolution models in 
section \ref{compare}; our conclusions follow in section \ref{concl}.
All the magnitudes are Vega relative unless otherwise specified.

\section{The Survey of a Wide Area with NACO} \label{SWAN}

Having already characterized large samples of objects in bright star 
fields, as described in the previous section, we targeted them with NACO on 
the VLT in order to exploit the present generation of AO technology for galaxy 
evolution studies.  NACO comprises the NAOS Shack-Hartmann AO module (Rousset 
et al. \cite{rousset}) mated with CONICA near-infrared camera (Lenzen et 
al. \cite{lenzen}).  Our choice of NACO observing mode was dictated by our 
desire to complement previous HST/NICMOS surveys.  First, we 
chose to image in $K_\mathrm{s}$, where NICMOS is less sensitive than in $J$ 
and $H$, thus making SWAN preferentially sensitive to red objects. Second, we 
chose to prioritize survey area over depth, in order to optimize the study 
of the galaxies over the last half of the Hubble time and improve SWAN's  
sensitivity to rare objects and its robustness against cosmic variance (the 
latter already enhanced by the survey's peculiarity of patching together small 
fields at different locations on the sky).  Use of NACO's $0.054\arcsec$ pixel 
scale (to maximize the field of view) and the Strehl ratios of 30--60\% typically 
achieved in $K_\mathrm{s}$ result in images that are slightly undersampled.  As 
the AO PSF is quickly changing both in time and position on the frame, in 
order to extract full information from our wide-field observations we have 
developed a new approach to account for the anisoplanatic PSF. The method was 
presented in Cresci et al. (\cite{cresci}), hereafter Paper I, along with some 
examples of galaxy morphology fitting using the derived model PSF.

\begin{figure}
    \begin{center}
     \resizebox{\hsize}{!}{\includegraphics[clip]{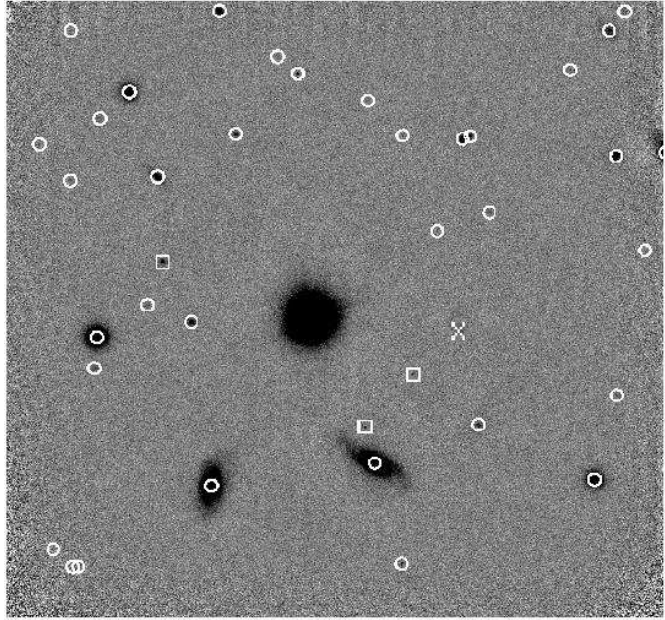}}   
    \caption{Example of a $55\arcsec \times 55\arcsec$ SWAN field: 
    SBSF\,24. The bright source in the center is the guide star, and 
    the circles are the extended 
    objects detected by SExtractor (SExtractor stellarity index 
    $\mathrm{SSI} < 0.9$); the squares are point sources ($\mathrm{SSI} 
    \geq 0.9$).  A ghost of the bright guide star is marked with a cross. 
    North is up and east is right.}
    \label{swanfield}
    \end{center}
\end{figure}

Each NACO pointing provides a usable $\sim 0.75\,\mathrm{arcmin}^2$ of the 
full $55.5\arcsec \times 55.5\arcsec$ detector area, due to losses from 
dithering and the central star (see, e.g., Fig.~\ref{swanfield}).  
Nevertheless, the anticipated survey area that will result from assembling 42 
such images will be -- at $\sim 30\,\mathrm{arcmin}^2$ -- some six times 
larger than the NICMOS survey of the HDF and flanking fields in $J$ and $H$ 
(Dickinson \cite{dickinson99}; Dickinson et al. \cite{dickinson00}).  
SWAN aims to combine the 
high angular resolution of a space-based survey with the shallower depth and 
wider area of a ground-based survey, thereby probing sources that are compact, 
faint, red, and rare more effectively than any other survey to date.

%
%
\begin{table*}

\begin{center}
\begin{tabular}{cccccccccc} 

\noalign{\smallskip} \hline \noalign{\smallskip}
%
%
\noalign{\smallskip} 
 Name & R.A. & DEC. & Obs. & $\Delta t$ & $\sigma_{\rm image}$$^b$ & mean & Strehl & N. of & $\theta_0$$^c$ \\

 & (J2000.0) & (J2000.0) & Date$^a$ & (min) & (e$^-$\,s$^{-1}$) & airmass & \% & Stars & (\arcsec) \\

(1) & (2) & (3) & (4) & (5) & (6) & (7) & (8) & (9) & (10)  \\

\noalign{\smallskip} \hline \noalign{\smallskip}

SBSF\,02  & 00 44 31.88 & $-$29 52 30.3 & 07.09.04   &   54  & 0.29 & 1.08 & --- & 2 & 5.90 \\
SBSF\,03  & 00 45 20.62 & $-$29 56 46.0 & 08.09.04   &   54  & 0.29 & 1.24 & --- & 2 & 11.20\\ 
SBSF\,04  & 00 45 28.05 & $-$29 31 40.1 & 09.12.04   &   68  & 0.28 & 1.22 & 33  & 2 & 10.65 \\ 
SBSF\,06  & 00 50 34.70 & $-$29 26 32.0 & 09.09.04   &   54  & 0.33 & 1.17 & 24  & 0 & [12.70] \\
SBSF\,08  & 00 52 18.88 & $-$29 27 17.8 & 12.09.04   &   56  & 0.32 & 1.09 & 42  & 1 & 12.00 \\
SBSF\,14  & 06 07 06.34 & $-$13 13 37.1 & 15.12.02   &   60  & 0.22 & 1.20 & --- & 4 & 11.95\\
SBSF\,15  & 08 44 00.22 & $-$16 34 01.1 & 17.12.02   &   60  & 0.25 & 1.12 & 46  & 10 & 21.78\\
SBSF\,16  & 09 14 52.77 & $-$19 26 17.0 & 18.12.02   &   50  & 0.25 & 1.07 & --- & 1 & 12.84\\
SBSF\,17  & 09 47 44.79 & $-$21 37 12.7 & 20.03.03   &   44  & 0.28 & 1.05 & 35  & 2 & 10.61\\
SBSF\,18  & 09 49 46.99 & $-$21 45 13.3 & 17.12.02   &   40  & 0.30 & 1.06 & --- & 7 & 21.18\\
SBSF\,24  & 10 40 26.20 & $-$30 00 36.5 & 21.03.03   &   60  & 0.24 & 1.02 & 30  & 8 & 11.77\\
SBSF\,27  & 12 55 37.48 & $-$31 42 41.3 & 11.04.04   &   44  & 0.30 & 1.07 & 35  & 12 & 17.85\\
SBSF\,28  & 12 56 14.14 & $-$42 09 10.9 & 11.04.04   &   44  & 0.27 & 1.26 & 42  & 2 & 9.94\\
SBSF\,34  & 13 46 25.24 & $-$31 45 47.5 & 11.04.04   &   44  & 0.28 & 1.36 & 28  & 2 & 4.89\\
SBSF\,36  & 22 14 36.74 & $-$28 25 31.6 & 05.09.03   &   60  & 0.23 & 1.09 & 35  & 2 & 13.25\\
SBSF\,37  & 22 43 04.43 & $-$39 49 29.3 & 06.09.03   &   60  & 0.23 & 1.10 & 34  & 0 & [12.70]\\
SBSF\,38  & 22 47 06.77 & $-$40 10 01.3 & 05.09.03   &   60  & 0.22 & 1.04 & 31  & 2 & 10.68\\
SBSF\,39  & 22 49 34.23 & $-$39 33 05.3 & 15.06.03   &   60  & 0.24 & 1.04 & 22  & 1 & 10.00\\
SBSF\,40  & 22 49 49.32 & $-$39 53 15.0 & 12.06.03   &   48  & 0.34 & 1.04 & 10  & 0 & 12.70\\
SBSF\,41  & 22 50 21.28 & $-$40 07 38.6 & 14.06.03   &   60  & 0.26 & 1.04 & 33  & 3 & 19.72 \\
SBSF\,42  & 23 29 55.77 & $-$18 35 54.1 & 16.06.03   &   60  & 0.24 & 1.03 & 35  & 0 & [12.70] \\

\noalign{\smallskip} \hline \noalign{\smallskip}

\end{tabular}
\end{center}
\caption{ Observational parameters and AO performances for NACO observations of SWAN fields. 
See the text for a full description of the entries. 
$^a$ CONICA was fitted with a new detector in June 2004.
$^b$ The noise is that measured in the resulting co-added image, scaled to 
a 1\,sec integration. Its statistical properties closely follow a Gaussian 
distribution with additional weak wings.
$^c$ `$\theta_0$' refers to the isoplanatic angle in $K_\mathrm{s}$ band 
as measured fitting the variation of the Strehl ratio of the point sources 
in the fields as described in the text. 
}
\label{obspar}

\end{table*}
%

\section{Observations} \label{obs}

The first 21 SWAN fields were observed in $K_\mathrm{s}$ band with the NACO AO 
system at the VLT, using the visible Wave Front Sensor (WFS). 
An example of a SWAN image is given in 
Fig.~\ref{swanfield}.  Table~\ref{obspar} summarizes other observational 
parameters and the AO system performance during the observations.  The SBSF 
name is given in column [1], and the coordinates of the guide star in the 
center of each field are given in [2] and [3], accurate to $\pm 0.2\arcsec$. 
Column [4] reports the date(s) on which each field was observed.
The total integration time on each field is given in [5], and 
the noise measured in the resulting coadded image rescaled to 1\,sec 
integration time in [6]. The mean airmass is reported in [7].
The Strehl ratio, estimated from a series of short exposures through a
narrow band filter taken before and/or after the science exposures in 
order to monitor the on-axis PSF, is given in column [8]. 
This is calculated from the ratio of the maximum pixel to the total
flux, and includes a correction for the offset of the PSF's centroid
from the center of a pixel; this can be considerable (typically
adding 5--10\% to the Strehl ratio for the data here) due to the large
0.054\arcsec\ size of the pixels. 
The number of bright point sources in each field used to evaluate 
the isoplanatic angle at 2.2\,${\rm \mu m}$, fitting the 
variation of their Strehl ratio in our $K_\mathrm{s}$ images 
(see section \ref{morph}), is reported in [9], and the 
resulting isoplanatic angle in column [10]. 

\section{Data reduction} \label{reduc}

The data obtained were reduced using
PC-IRAF (version~2.11.3) together with some scripts in IDL (version~6.0).
The presence of a bright star in the center of a field less than
1\arcmin\ across made the data reduction more complex than usual,
requiring extra steps to compensate.
An initial estimate of the sky background was made from the target
frames after masking all bright objects in the fields.
Each target frame then had the sky subtracted, was flat fielded, had any
residual constant background removed, and hot pixels corrected.
A mask that included dead pixels and bad regions 
was then applied to each frame.
In order to correct for over-subtraction from very extended faint
scattering (and/or emission) around bright objects, a surface was fit
to each frame (ignoring regions in the object mask) and subtracted.
The frames were then aligned with sub-pixel accuracy using up to
several conspicuous isolated objects in the field, and averaged after
rejecting high and low pixels at each point according to an estimated
variance.
This initial combined frame was used to generate a new object mask, and 
the entire data reduction process was repeated, 
yielding a new combined frame with much less over-subtraction.
In a final step, the objects were once more masked out and a surface
fitted to the background, and this was subtracted to produce the final
image.

The sky is estimated by dithering, i.e., slightly moving the telescope 
between different frames so that different pixels sample different parts 
of the sky. In SWAN the offsets were chosen semi-randomly within a 7" box, due 
to the limited NACO field of view. Therefore, even if great care is used 
to produce the sky frames, the sky around objects larger than $\sim 3\arcsec$,
i.e., for the very bright guide star and for galaxies with effective radius 
$R_e \gtrsim 1\arcsec$, may be overestimated, producing a 
self-subtraction of some galaxy flux. This effect can produce fainter 
magnitudes and smaller dimensions for such bright and large objects, 
although these constitute less than 2\% of the total sample in our fields. 
However, in section \ref{compare} we will see that this effect introduces 
some systematic uncertainties in the size-magnitude relation for large galaxies.

\section{Source detection} \label{detect}

In each reduced SWAN field, sources were detected using SExtractor 
(Bertin \& Arnouts \cite{bertin}), with the
appropriate parameters optimized for compact sources, 
set to provide a positive detection for objects
brighter than $1.5\,\sigma$ per pixel over an area of more than
3~pixels. To improve the detection of faint sources we used a Gaussian 
filter ($\sigma=1.5$~pixels) to smooth the image. 
False detections at the noisy borders of the mosaic and on the spikes 
and the ghost of the bright guide star were removed.  For the former, a mask 
that indicated the fraction of the total integration time spent on each pixel 
was used; objects detected in pixels below a specified threshold were rejected.
For the latter, appropriate object masks were created.  Our algorithm 
deliberately does not push the detection to the faintest possible limit, as we 
are more interested in the high resolution AO morphologies of the brighter 
sources than in the deepest possible number counts.  For this reason, our 
counts (see Section \ref{counts}) are not significantly contaminated by 
spurious detections due to noise.  The total coverage above the detection 
thresholds of the 21 fields is $15.3\,\mathrm{arcmin}^2$, within
which a total of 495 sources are detected down to a magnitude of
$K_\mathrm{s} \sim 23.5$ ($K_{\mathrm{AB}}\sim 25.3$, see section \ref{counts}).

\section{Morphological fitting} \label{morph}

The morphological parameters of the detected galaxies were derived using 
GALFIT (Peng et al. \cite{peng}), a widely used software package that fits 
a two-dimensional image of a galaxy and/or a point source with one or 
more analytic functions that have been convolved with a model of the PSF. 
To fit the galaxies in our SWAN fields we used  
a single S\'ersic (\cite{sersic}) profile,
\begin{equation}
	I(R) = I(R_e) \times \exp ({-b_n \times [(R/R_e)^{1/n}-1]})
\end{equation} 
where $R_e$ is the effective radius that encloses half of the light, 
$n$ is the S\'ersic index and $b_n$ is a constant that varies with $n$, chosen 
so that $R_e$ corresponds to the half-light radius.
 
GALFIT needs as an input a PSF to convolve the S\'ersic profile 
model. We used the off-axis AO PSF model presented in Paper I, which is 
optimized for wide-field and high Galactic latitude observations. 
The off-axis PSF is determined by convolving the on-axis PSF 
in each of the fields with an elliptical Gaussian kernel elongated 
towards the guide star. The FWHM of the kernel depends on the distance from 
the guide star and on the isoplanatic angle of the field. 
We therefore derived the isoplanatic angle for each field 
fitting the variation of the Strehl ratio of the point sources across the 
field as described in Paper I. The obtained isoplanatic angle along with the 
number of point sources used in the fit are reported in Table~\ref{obspar}. 
The derived isoplanatic angles for the 21 fields range from 
$4.9\arcsec$ to $21.8\arcsec$. In four of the fields no bright point source 
was available except the guide star, and therefore the average isoplanatic 
angle for the other fields ($12.7\arcsec$) was assumed.

Initial guesses for GALFIT model parameters were 
obtained from the SExtractor source catalogs. Lacking an estimate of the 
S\'ersic index $n$ in the SExtractor catalogs, we used $n=2$ for all the 
galaxies in the first iteration. Each galaxy was fitted twice, 
using as first guesses for the second iteration the output parameters of the 
first iteration.  Roughly 16\% of the detected galaxies could not be fitted 
satisfactorily with a single component, but required simultaneous fits with 
very close companions or multiple-component fits.  These can be divided in two 
categories. 9\% of the total are 
interacting galaxies or very close pairs, where the overlap of the isophotes 
from different objects required a simultaneous fit.  A further 7\% of the 
total are galaxies for which a single-component S\'ersic profile was not 
sufficient to fit the light profile, leaving significant residuals in the 
subtraction.  Half of these two-component galaxies were re-fit using a 
disk component and an elliptical bulge, while the other half were re-fit by 
adding a central point source to the S\'ersic component.

As we have shown by the detailed simulations in Paper I, the morphological 
parameters of the galaxies detected at the depths of our images 
can be derived with low uncertainties up to $K_\mathrm{s} \sim 20.5$, 
while for fainter objects the 
uncertainties grow as a function of the magnitude. In addition, we recall that
it is possible to set a threshold of $n = 2$ on the S\'ersic index that can 
discriminate between late-type galaxies ($n<2$) and early-type galaxies 
($n>2$).  The results of our simulation are confirmed e.g. by Ravindranath et 
al. (\cite{ravindra04}), who used GALFIT to fit single S\'ersic profiles to a 
sample of nearby galaxies of known morphology from the Frei et al. 
(\cite{frei96}) sample, after artificially redshifting them to $z=0.5$ and 
$z=1.0$. They found that $n=2$ is the appropriate threshold to separate 
disk-dominated galaxies from bulge-dominated ones, even in the presence of 
morphological complexities such as dust, star-forming regions, etc. 
(Ravindranath et al. \cite{ravindra04}).

Of the 383 galaxies detected to $K_\mathrm{s} \sim 23.5$ (see section 
\ref{stargal} for a discussion of the 112 stars), 214 were classified 
as late-type and 169 as early-type. The sources fitted with multiple 
components are classified according to the S\'ersic component providing the 
higher flux contribution. The galaxies are divided in 
these two subclasses for the following analysis, with an average contamination 
between the two subclasses of less than 10\% up to $K_\mathrm{s}=21$ (Paper I).
\begin{figure}
        \begin{center}
        \resizebox{0.4\textwidth}{!}{\includegraphics{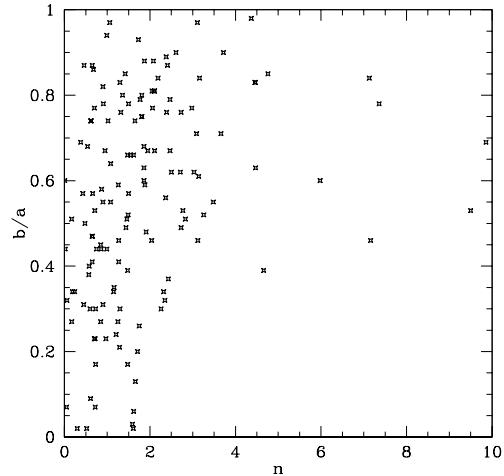}}  
        \caption{\label{ellitt} The distribution of the axis ratios 
	$b/a$ of the SWAN galaxies fitted by GALFIT with $\chi^2_{\nu} \leq2$ 
	as a function of their S\'ersic index $n$. 
	As expected, while late-type galaxies ($n<2$) are observed at random 
	inclinations with respect to the plane of the sky, 
	and therefore at every $b/a$, early-type galaxies ($n>2$) 
	are not observed with $b/a \lesssim 0.4$.}
     \end{center}
\end{figure}
In order to quantify the morphological fit quality, we used 
the $\chi^2_{\nu}$ calculated by GALFIT. We classified as well-fit the
315 galaxies with $\chi^2_{\nu} \leq2$ (167 late-type and 148 early-type), 
while the other fits were considered less reliable and are not considered when computing 
the size-magnitude relation of the galaxies in the SWAN fields 
(although they are included in the number counts).  As an
additional check of our late/early-type separation, we show in 
Fig.~\ref{ellitt} the distribution of the axis ratios $b/a$ of the 
galaxies with $\chi^2_{\nu} \leq2$ as a function of S\'ersic
index $n$. As expected, while the late-type galaxies are observed at random 
inclinations with respect to the plane of the sky, and therefore at
every $b/a$, early-type galaxies are not observed with $b/a \lesssim
0.4$ (e.g. Lambas et al. \cite{lambas92}). This confirms that our 
morphological classification of early and late type galaxies based on
the S\'ersic index $n$ produces reliable results.

While the redshifts of these objects are presently unknown, 
the magnitude-redshift relation of Cowie et al. (\cite{cowie})
and the K20 survey (Cimatti et al. \cite{cimatti}) 
indicate that at $K = 20$ the median redshift is $z \sim 0.8-1$.
At this redshift, our spatial resolution of $0.1\arcsec$, which also
corresponds to the smallest effective radius bin, is equivalent to
only 500\,pc for typical cosmologies, hinting at the exciting
potential of this work.

\section{Star-galaxy separation} \label{stargal}

The separation of Galactic foreground stars from the field galaxies 
is a critical step for avoiding star contamination in our galaxy catalogue. 
We classified as stars all 58 sources detected in the NACO images 
with SExtractor stellarity index $\mathrm{SSI} \geq 0.9$. 
The SExtractor classification should be treated with caution
since it assumes a constant PSF across each field, and elongated sources 
are more likely classified as galaxies. However, 
all the objects classified as stars by SExtractor lie on an upper 
envelope in a Strehl versus radial distance plot, i.e., they have the 
highest Strehl ratio among the sources at the same distance from 
the guide star, supporting their classification as point sources. 
It remains possible that some stars are not classified 
as point sources by SExtractor, due to the limited isoplanatic AO 
patch and their resulting elongated shape. We therefore also 
classified as stars all the very compact sources fitted 
by GALFIT with $R_e<0.01\arcsec$. 
This is supported by simulations in which we fitted true 
fiducial point sources in the SWAN fields, rescaled to several 
magnitudes, with GALFIT S\'ersic profiles and obtained 
very compact effective radii $R_e<0.01\arcsec$ and high S\'ersic indexes.  
For very bright and elongated PSFs, the fitted $R_e$ can still  
be as large as $\sim 0.2\arcsec$, due to the higher 
signal in the halos of the PSF that may not be perfectly 
reproduced by the PSF model. We therefore include in the star 
catalogue all the sources with $K_\mathrm{s} \leq 18.5$ (in order 
to have sufficiently high S/N) that are classified as 
stars by SExtractor in our SOFI seeing limited images of the 
SWAN fields. All the objects classified as stars in the seeing-limited images 
proved to be compact in the AO-corrected ones as well, even if elongated, 
with all having $R_e < 0.16\arcsec$ as fitted by GALFIT 
using the appropriate local PSFs for convolution.

\begin{figure}
        \begin{center}
	\resizebox{0.4\textwidth}{!}{\includegraphics{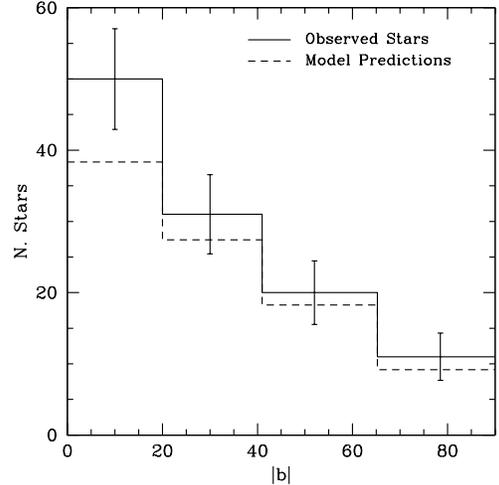}}
	\caption{\label{starsswan} Number of stars in the SWAN fields as a function 
	of Galactic latitude $b$ up to $K_\mathrm{s} = 22$, compared with the 
        predictions of 
	the Galaxy model of Bahcall et al. (1980). 
	The observed and predicted stars number counts are 
	in very good agreement for all latitudes except the lowest latitude 
        bin ($|b|<20^\circ$), 
	where the Galactic model is less accurate due to the high 
	variability between adjacent lines of sight.}
	\end{center}
\end{figure}

We have a total of 112 stars in the 21 SWAN fields analyzed. 
To assess the robustness of the star/galaxy classification, 
the star counts were compared with the predictions of the Bahcall et al. 
(\cite{bachall80}) galaxy model, which provides the 
star counts as a function of the field's Galactic longitude and latitude. 
As the model provides the number of stars brighter than a certain limiting 
magnitude in the $V$ band, we convert the $V$ magnitude into a $K_\mathrm{s}$ 
magnitude using an average color derived from the $K$-band counts at 
the Galactic pole provided by Hutchings et al.   
(\cite{hutchings02}). In Fig.~\ref{starsswan} we show the number of stars 
in the SWAN fields as a function of Galactic latitude $b$ up to $K_\mathrm{s} 
= 22$, which 
corresponds to the limit where we are 100\% complete for point sources 
(see Fig.~\ref{corrps}). It can be seen that the observed and predicted stellar number counts 
are in very good agreement for all latitudes except the lowest latitude bin 
($|b|<20^\circ$), where the Galactic model is less accurate due to the high 
variability between adjacent lines of sight. However, the total excess of 
selected stars with respect to the model predictions is only 18 sources, i.e., 
small compared to the total catalogue of 383 galaxies. Therefore, even if some 
compact galaxies in these fields were erroneously classified as stars, 
they represent less than 5\% of the sample.

\section{Completeness correction and number counts} \label{counts}

The probability of detecting a source in one of our images depends
on five different parameters:
\begin{figure*}
        \centering
        \begin{minipage}[c]{1.0\textwidth}
        \centerline{\hbox{
                  \psfig{file=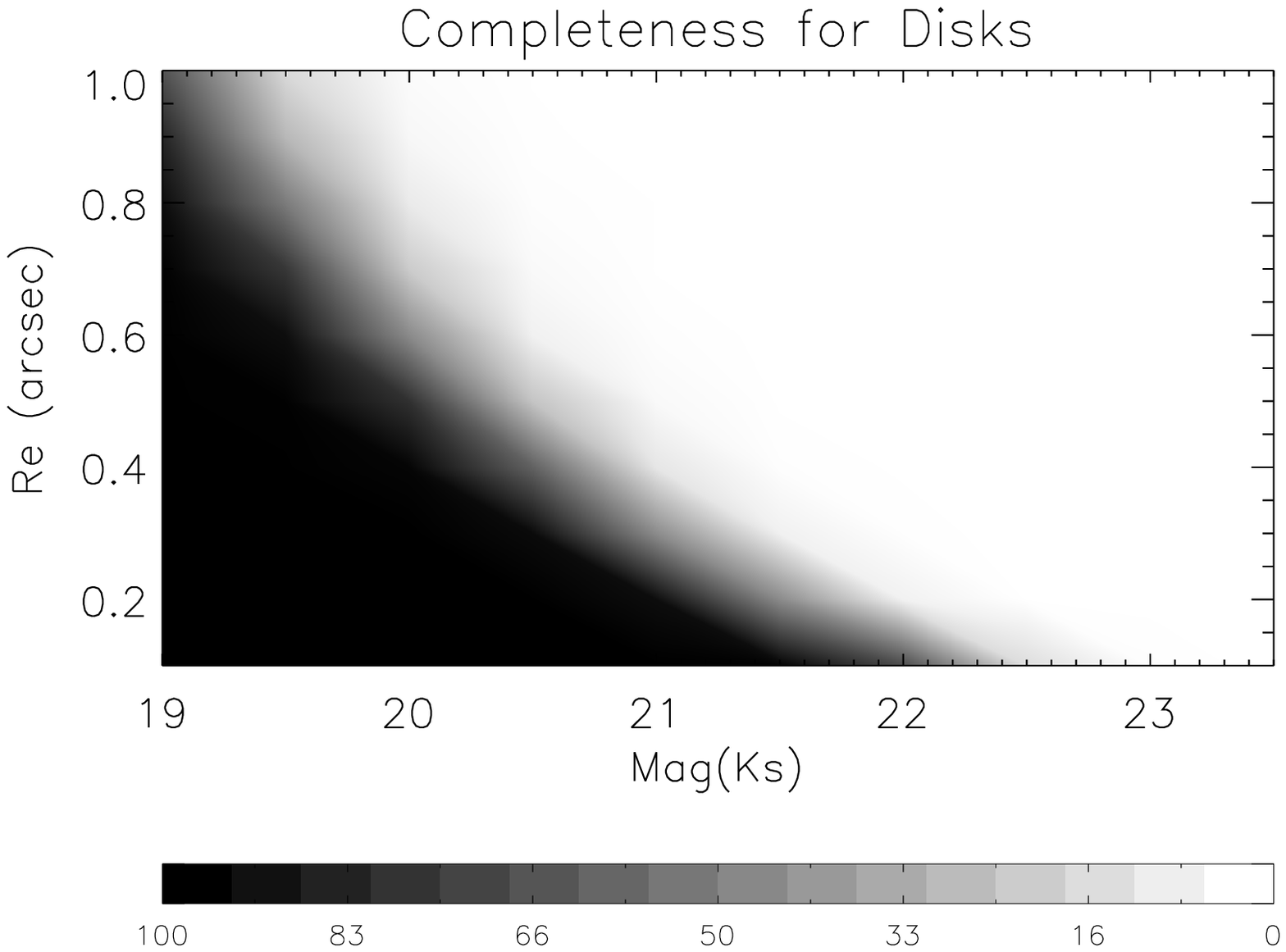,width=9.0cm}
                  \hspace{0.0cm}
                  \psfig{file=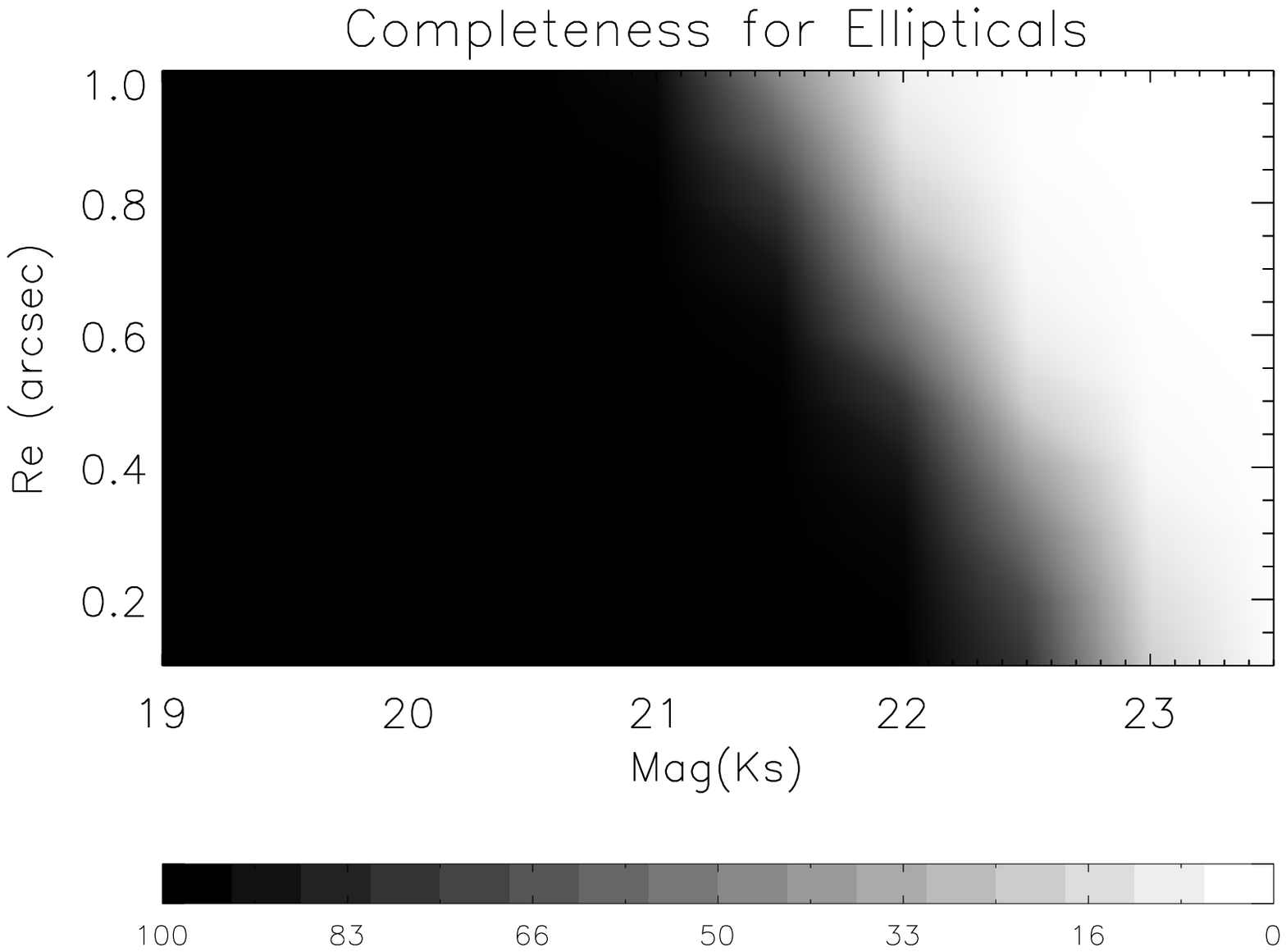,width=9.0cm}
        }}
        \caption{The \textit{left panel} shows the variation of the detection 
	probability for a late-type galaxy in SWAN as a function of the
        magnitude and the effective radius 
	$R_e$. The probability is the average for all 21 observed 
	fields, and assumes a distance from the guide star 
	$1 < \theta/\theta_0 \le 2$. The \textit{right panel} shows
        the same for early-type galaxies.}  
        \label{compl}
        \end{minipage}
\end{figure*}
\begin{enumerate}
\item The total integrated magnitude
\item The S\'ersic index $n$, as for a given magnitude more
concentrated objects (i.e., early-like galaxies with $n >2$) 
are more easily detected than exponential-like galaxies ($n<2$) with 
lower concentration.
\item The effective radius $R_e$. As before, 
more compact sources are more easily detected.
\item The SWAN field in which the source was observed. The 
integration time and therefore the signal/noise ratio is different in  
different fields. In addition the overall AO correction is different in each 
observation, as is indicated by the different on-axis Strehl ratios (see 
Table~\ref{obspar}). 
\item The distance from the guide star $\theta/\theta_0$, as the
degree of correction of the AO system depends strongly on this
parameter (see, e.g., Paper I).
\end{enumerate}
The last two parameters are due to the distinctive attributes of our survey, 
which makes use of several different fields (4) and of AO (4,5).

\begin{figure}
    \begin{center}
     \resizebox{0.4\textwidth}{!}{\includegraphics{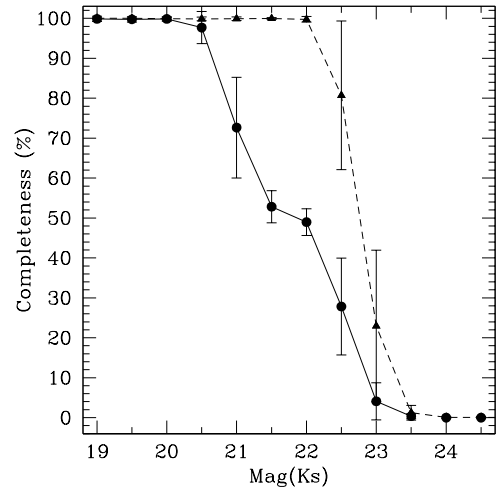}}   
     \caption{Comparison between the completeness for point sources and for 
     extended sources for the SWAN fields. The completeness for point sources 
     (triangles, dashed line) 
     was evaluated adding 100 true NACO point sources ($\theta/\theta_0=1.5$) 
     for each magnitude to each field and then averaging over all fields. The 
     completeness for extended sources (circles, solid line) 
     is the average over all the fields for both 
     late and early-type with $R_e=0.3\arcsec$, i.e., the mean for all the 
     detected sources in SWAN, at the same distance from the guide star used 
     for the point sources. The errorbars show the variance for the 21 
     different fields.} 
     \label{corrps}
     \end{center}
\end{figure}

In order to derive the detection probability for each combination of
these five parameters, we ran several simulations, adding a total of
65,000 simulated galaxy profiles with known parameters -- matched to
the ones of the observed galaxies -- to the original SWAN fields at random 
locations and tried to recover them running SExtractor again. 
We used extended sources to evaluate the 
completeness correction, as this produces results that are quite different from
those inferred using point sources alone, especially at this resolution.
SExtractor was used with the same parameters used in the science
source detection.  We used the S\'ersic index $n=1$ for late-type galaxies and 
$n=4$ for early-type galaxies. For both types the effective radius $R_e$ 
ranged from 0.1\arcsec to 1.0\arcsec. The galaxy profiles were
convolved with real NACO PSFs extracted from point sources in our data 
lying at different distances from the guide star, in order to simulate
the effect of the AO correction.
The simulated galaxies have magnitudes ranging between $K_\mathrm{s}=19$, where 
we are 100\% complete for every combination of the other four parameters, 
to $K_\mathrm{s}=23.5$.  
We consider three different regimes for the detection probability 
as a function of the distance from the guide star: $\theta/\theta_0 \le 1$, 
$1<\theta/\theta_0 \le 2$ and $\theta/\theta_0 >2$. 
We used point sources at $\theta/\theta_0=$ 0.9, 1.5, and 2.8 respectively 
for the three regimes as references for the PSF in the simulated galaxies.

Using this approach, we can derive the detection probability for a
galaxy of known magnitude, $R_e$, S\'ersic index, and distance from
the guide star in a particular field.  By way of example, the
histograms of the detection probability averaged over all 21 fields as
a function of magnitude and of effective radius $R_e$ for late and early-type
galaxies with $1 < \theta/\theta_0  \le 2$ are shown in  
Fig.~\ref{compl}. It is clear from comparing the panels how much more 
sensitive high resolution images are to more core-concentrated sources 
like the elliptical galaxies.

The detection probability can be used to correct the number counts
using the observed galaxy population as a starting point. From our simulations 
we derived the detection probability $P_{\rm detect}$ for
each detected galaxy in the survey, using the measured $R_e$,
magnitude, and S\'ersic index from GALFIT fitting (see section 
\ref{morph}), along with the position of the galaxy in the field. 
Each galaxy is then regarded in the completeness-corrected number counts 
as $1/P_{\rm detect}$ sources at its magnitude, so that, for example, 
a galaxy with $P_{\rm detect}=0.80$ counts as 1.25 galaxies once the
correction is applied. 

In Fig.~\ref{corrps} we show the average completeness over all the fields 
for both point sources and extended objects as a function of
magnitude.  The completeness for point sources was  
derived by adding (100 times) to each field a true NACO point source with rescaled
flux.  The point source was at a distance
$\theta/\theta_0 = 1.5$ from the guide star.  
The completeness for extended objects is the average between that of late-type 
and of early-type galaxies 
over all the fields, using the same NACO point source PSF to convolve the 
simulated galaxy profiles. The effective radius was fixed at the
average for the detected SWAN sources ($R_e=0.3\arcsec$). Obviously
the correction derived using only point sources is much smaller than the
one derived as described above, with the number of sources in the
range $20.5 \le K_\mathrm{s} \le 22$ (where no correction would be
applied in the point-source case) being particularly underestimated.

\begin{figure}
    \begin{center}
     \resizebox{\hsize}{!}{\includegraphics{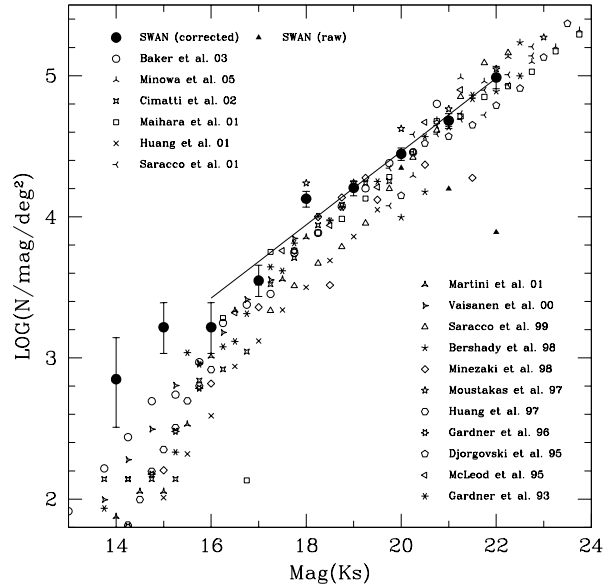}}   
    \caption{$K_\mathrm{s}$ corrected (solid circles) and uncorrected
      (solid triangles) number counts from SWAN, compared with other
      $K$-band surveys. The points
      at $K_\mathrm{s} \ge 23$ are not reliable as the completeness
      correction is adding more than 90\% of the sources.  
      The excess at bright magnitudes is due to a selection bias, as
      some of the fields were selected to contain more bright galaxies
      (Baker et al. 2003). The solid line is the best fitting
      power-law in the range $16 \le K_\mathrm{s} \le 22$,  
      with a slope $d\,
	\mathrm{log}\,(N)/dM=0.26$. The error bars show the
      Poissonian errors.} 
    \label{corrcounts}
    \end{center}
\end{figure}

The corrected number counts, obtained using the detection probabilities
of the observed SWAN sources as weights, are shown in Fig.~\ref{corrcounts}.
\begin{table}
	\begin{center}
	\begin{tabular}{ccccc}
	\noalign{\smallskip} \hline \noalign{\smallskip}
	Mag($K_s$) & $n_\mathrm{raw}$ & $N_\mathrm{tot}$ & Late-type 
& Early-type \\
	(1)	   & (2) & (3) & (4) & (5) \\
	\noalign{\smallskip} \hline \noalign{\smallskip}
	  14    &      3   &       7e+02 &      2e+02  &    5e+02 \\
          15    &      8   &        1.9e+03 &      5e+02  &     1.4e+03 \\
          16    &      8   &        1.8e+03 &       1.4e+03  &    4e+02 \\
          17    &     19   &        4e+03 &       1.4e+03  &     3.0e+03 \\
          18    &     62   &   1.5e+04 &       4.9e+03  &     9e+03 \\
          19    &     69   &   1.6e+04 &       9e+03  &     7.8e+03 \\
          20    &     96   &    2.8e+04 &  2.0e+04  &     8.7e+03 \\
          21    &     67   &   4.8e+04 &   4.1e+04  &     6.8e+03 \\
          22    &     35   &   9.8+04 &  9.3e+04  &     4.4e+04 \\
	\noalign{\smallskip} \hline \noalign{\smallskip}
	\end{tabular}
	\caption{\label{tabcounts} Differential number counts in the
        SWAN fields.  
	The raw number of detected galaxies is reported in (2). In (3) 
	the corrected number counts 
	($\textrm{N}\ \textrm{mag}^{-1}\ \textrm{deg}^{-2}$)
        for the whole sample are shown, while (4) and (5) separate 
	late-type and early-type counts, respectively.}
	\end{center}
\end{table}
At brighter magnitudes ($K_\mathrm{s} \le 16$), an excess is present
in our number counts with respect to others in the literature. This is
due to a selection bias, as some of the fields were chosen to have an
excess of bright galaxies (see Baker et al. \cite{baker}). At fainter 
magnitudes, the points at $K_\mathrm{s} \ge 23$ are no longer
reliable, as the completeness correction adds more than 90\% of
the sources. We derive a slope $d\,{\rm log}\,(N)/dM=0.26 \pm 0.01$ in
the range $16 \le K_\mathrm{s} \le 22$. Our corrected number counts
are in good agreement with those found by other authors in the
literature (see Fig.~\ref{corrcounts}). 
This result, together with what was found in Paper I, shows once more
that we are able to account for the biases introduced by the variable
AO PSF and that the data can therefore be reliably used for further analysis.

\section{Comparison with galaxy evolution models} \label{compare}

Most theoretical models of galaxy formation and evolution can be
roughly divided into two broad categories: the so-called ``backwards'' 
approach and the ``{\it ab initio}'' approach.  In the former approach 
(Tinsley \cite{tinsley80}; Yoshii \& Takahara \cite{yoshii88};
Fukugita et al. \cite{fukagita90}; Rocca-Volmerange \& Guiderdoni 
\cite{rocca90}; Yoshii \& Peterson \cite{yoshii95}; Pozzetti et al. 
\cite{pozzetti96}, \cite{pozzetti98}; Jimenez \& Kashlinsky 
\cite{jimenez99}; Totani \& Yoshii \cite{totani00}; Totani et al. 
\cite{totani01}), the evolution is probed backwards into the past to 
predict observables such as galaxy counts and redshift distributions. 
The local properties of galaxies, like multi-band colors and chemical 
properties, are used to construct a reasonable model of the star formation 
history and luminosity evolution of galaxies based on the stellar
population synthesis method.  The formation epoch and merging history
of galaxies, however, cannot be predicted in this framework, as they
are introduced as phenomenological parameters that can be inferred
from comparison with observational data.

In the latter approach (Kauffmann et al. \cite{kauff93}; Cole et al. 
\cite{cole94}, \cite{cole00}; Somerville \& Primack \cite{som99}; 
Nagashima et al. \cite{naga05}), on the other hand,
the formation epoch and merging history of galaxies are predicted
by the standard theory of structure formation in the CDM universe.
In these models the local and high redshift properties of the galaxies 
such as the luminosity function, mass and size distribution, 
are outputs of the model that can be compared with observations. 
However, although the formation and evolution of dark matter halos are 
considered to be rather well understood, our knowledge about 
baryonic processes such as star formation, supernova feedback, and galaxy 
merging is still very poor, and a number of phenomenological parameters must 
be taken into account, making the comparison of the modeled and observed data 
more complex. Here we compare the galaxy counts in SWAN and 
the size-magnitude relation of the detected galaxies with
representative results of these two radically different approaches.

\subsection{The galaxy evolution models}

The first model used is a ``backwards'' pure luminosity evolution (PLE) model
developed by Totani \& Yoshii (\cite{totani00}) and Totani et al. 
(\cite{totani01}), based on the present-day properties of galaxies and
their observed luminosity function. It evolves a system's luminosity
and spectral energy distribution evolution backward in time using a 
standard galaxy evolution model in which star formation is tuned to 
reproduce galaxies' present-day colors and chemical properties
(Arimoto \& Yoshii \cite{arimoto87}; Arimoto et al. \cite{arimoto92}).
The model includes the effects of both internal dust obscuration and 
intergalactic H\,I absorption, and it does not incorporate galaxy mergers;
therefore the galaxy sizes and comoving number density do not evolve.
The number density of galaxies is normalized at $z=0$ using the local 
$B$-band 
luminosity function of galaxies, while the relation of the present $B$ 
luminosity and effective radius $R_e$ is determined from power-law fits 
to the empirical relation observed for local galaxies of different types 
(Bender et al. \cite{bender}; Impey et al. \cite{impey}).
Galaxies are in fact classified into six morphological types: 
three of them (Sab, Sbc, and Scd) are assigned to spiral galaxies,
while an Sdm model is used for irregular galaxies. Following Totani et al. 
(\cite{totani01}), we divided the E/S0 galaxies into distinct 
population of giant ellipticals (gE, $M_B \lesssim -17$) and dwarf 
ellipticals (dE, $M_B \gtrsim -17$). It is known that these are 
two distinct populations, showing different luminosity profiles 
(the $r^{1/4}$ law for giants and exponential for dwarfs; see 
Barazza et al. \cite{barazza05}), different luminosity-size relations, 
luminosity functions 
and different physical processes that govern the evolution of each type 
(see, e.g., Ferguson \& Binggeli \cite{ferguson94} and references therein; 
Ilbert et al. \cite{ilbert06} for evidence of two different populations 
up to $z\sim1.2$). 

In the 
$K_\mathrm{s}$ band, the critical separation magnitude ($M_B=-17$) 
corresponds to $M_{K_\mathrm{s}} \sim -21$ for 
the typical color of elliptical galaxies. Since the 
contribution of early-type galaxies is more significant in the 
near-infrared than in the optical, it is important to take into account 
such distinct populations of elliptical galaxies in predicting the 
$K_\mathrm{s}$ counts.

In addition, we compare the derived properties of the SWAN galaxies with the 
predictions of the ``\textit{ab initio}'' Numerical Galaxy Catalog (NuGC) 
of Nagashima et al. (\cite{naga05}), which is based on a semi-analytic (SA) model of galaxy 
formation combined with high-resolution $N$-body simulations in a 
$\Lambda$CDM cosmological framework.  The model includes several 
essential ingredients for galaxy formation, such as the merging histories 
of dark halos directly derived from $N$-body simulations, radiative 
gas cooling, star formation, supernova feedback, mergers of galaxies, 
population synthesis, and extinction by internal dust and intervening 
H\,I clouds.  The high resolution used for the simulations, with a 
minimum mass for dark halos of $3 \times 10^9\ M_{\odot}$, is
sufficient to resolve their effective Jeans mass. 

It was shown by Nagashima et al. (\cite{naga05}) that this model is in 
reasonable agreement with several observational results, like the 
luminosity functions in $B$ and $K$ bands, the H\,I mass function,
the size-magnitude relations for local spirals and elliptical
galaxies, the Tully-Fisher and Faber-Jackson relations at $z=0$, faint 
galaxy number counts in $BVRi'z'$ bands, and the cosmic star formation 
history at high redshift. In addition, the model is able to reproduce
the distribution of $(R-K)_{AB}$ colors with redshift observed in 
GOODS (Somerville et al. \cite{some04}), including extremely red 
($(R-Ks)_{AB}\sim 5$) galaxies that other semi-analytic treatments
have trouble accounting for.  In summary, the model is able to
reproduce several observational results for local and high-redshift 
galaxies, not just those that were used to tune the model parameters.

\subsection{Addressing cosmic variance} \label{cosmvar}

The uncertainties in galaxy number counts include contributions from 
Poisson errors and from the so-called ``cosmic variance'', due to the 
fact that galaxies are strongly clustered and thus distributed in
overdensities and large voids on the sky. 
Therefore, we have to take into account the corresponding effects 
on the relative normalizations of the predicted and 
observed counts in order to make a fair comparison. 
\begin{figure*}
        \centering
        \begin{minipage}[c]{1.0\textwidth}
        \centerline{\hbox{
                  \psfig{file=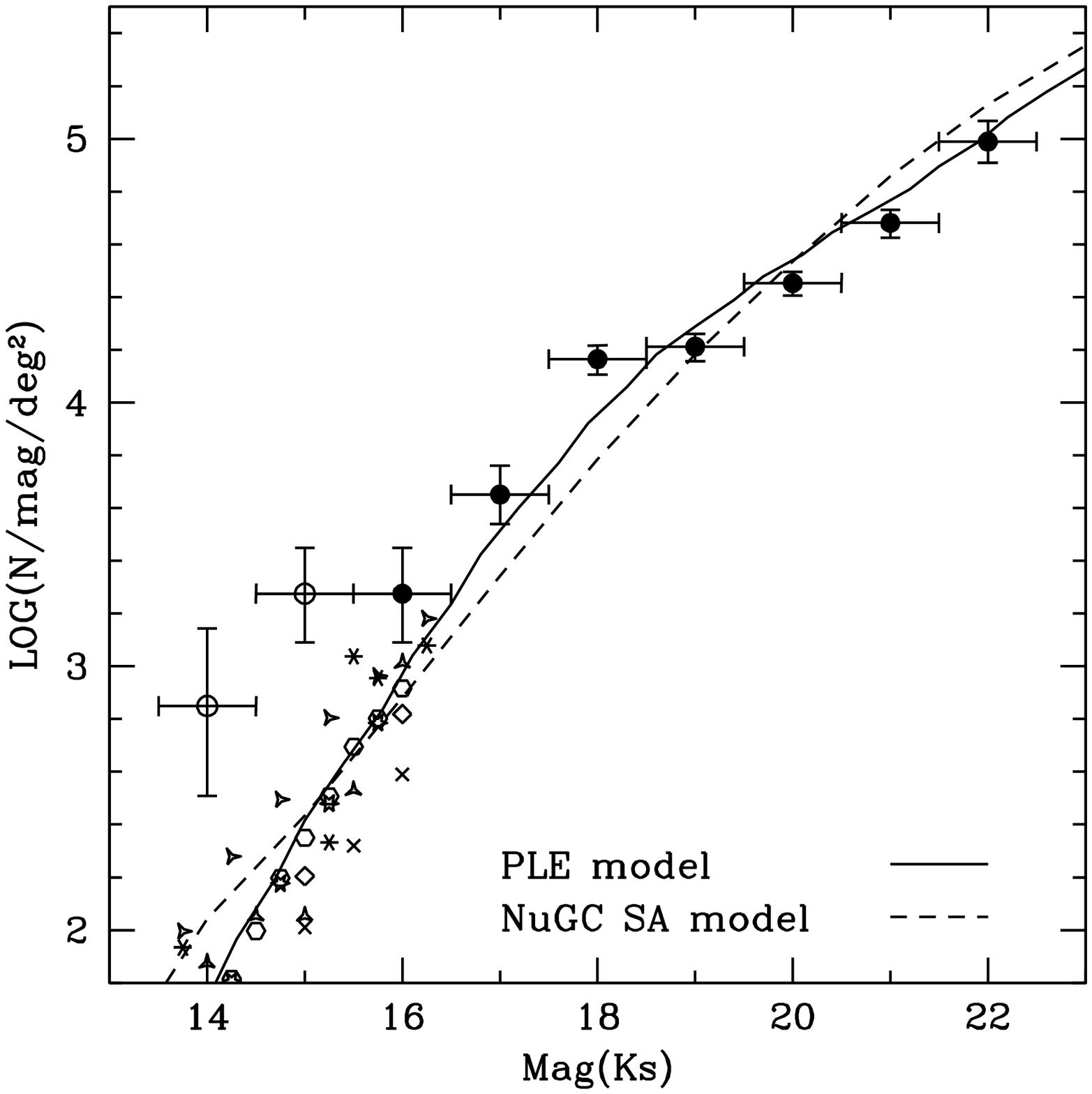,width=7.5cm}
                 \hspace{0.0cm}
                  \psfig{file=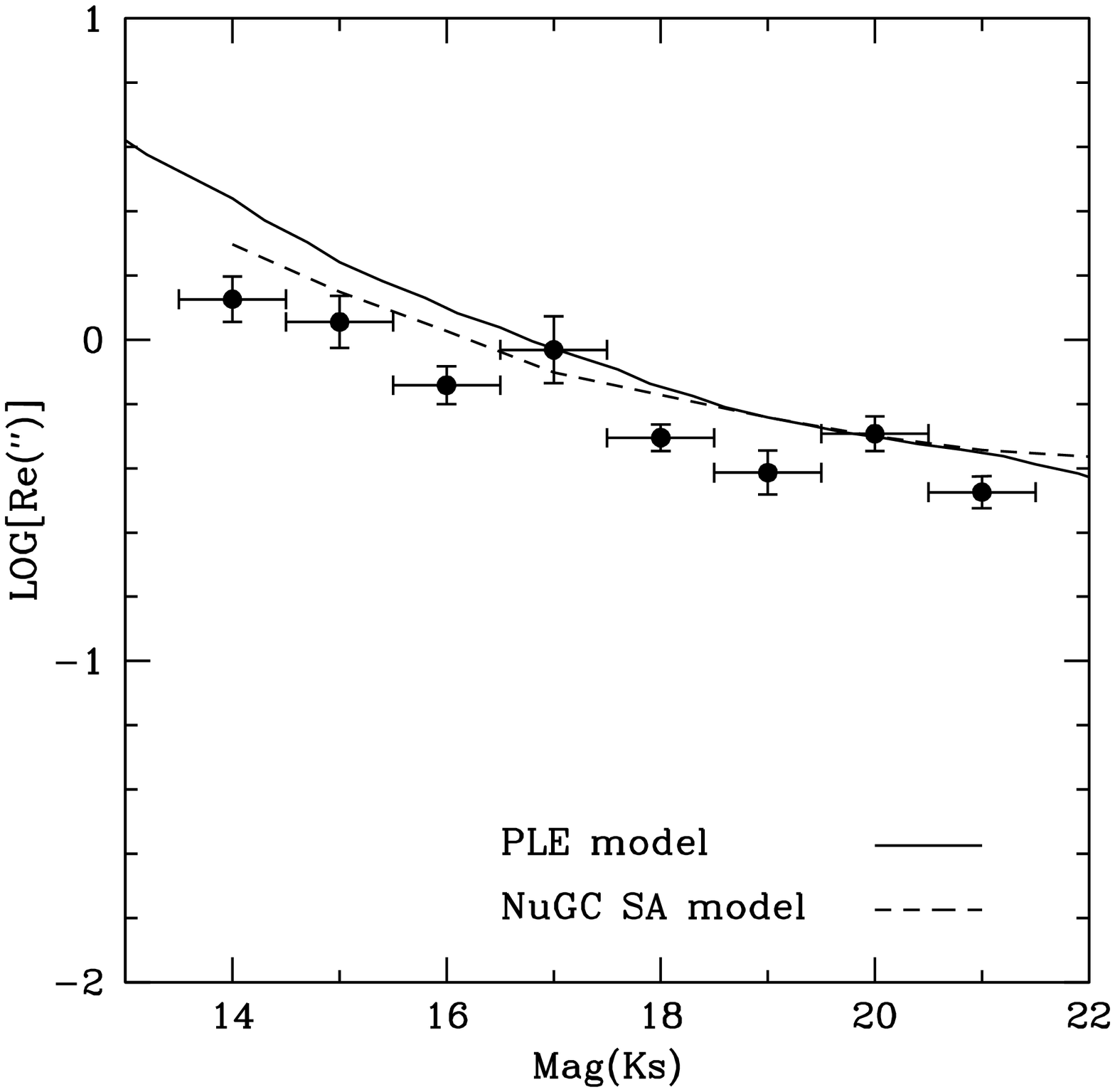,width=7.5cm}
        }}
        \caption{\label{tot} Comparison of total number 
	counts for all galaxies (\textit{left panel}), and of the mean 
	$R_e$ as a function of magnitude for those sources 
	with $\chi^2_{\nu} \leq2$ in the GALFIT fit (\textit{right panel}), 
	with the PLE model (solid line) and the hierarchical model (dashed line) 
        predictions.
	In the number counts, the points at $K_\mathrm{s} \leq 15$ 
        shows an excess due to a selection 
	bias, as some fields were selected to contain more bright galaxies. 
	Counts from other surveys in the literature for $K_\mathrm{s} \leq16$ 
	were therefore plotted for comparison to the model in this range 
	(see Fig.~\ref{corrcounts} for the references of the data points).
	The error bars on the number counts are the Poisson error,
	while in the size-magnitude relation they represent the
        standard error on the mean.}  
        \end{minipage}
\end{figure*}
According to Daddi et al. (\cite{daddi00}), the RMS fluctuations 
$\sigma_{tot}$ of numbers counts around their mean mean value, taking
into account both the Poisson error and the cosmic variance, are given by 
\begin{equation} \label{clust}
	\sigma^2_{tot}=X \cdot (1+ X\ AC)
\end{equation}
where $X$ are the total number counts, $A$ is the clustering amplitude 
($A \sim 1.6 \times 10^{-3}$ 
for the Daddi et al. (\cite{daddi00}) $K$-selected sample), 
and $C$ is a factor that depends on the 
field area and can be approximated by $C=58\,\mathrm{area}^{-0.4}$ for
an area expressed in arcmin$^2$. 
In our case, as we observe galaxies in $N = 21$ different fields
of the same area, thus reducing the clustering uncertainty, 
equation \ref{clust} can be written as 
\begin{equation}
	\sigma^2_{tot} \simeq X_{tot} \cdot \left(1+ \frac{X_{tot}}{N} AC\right)
\end{equation}
where $N$ is the number of fields, 
$X_{tot}$ is the total number of sources detected 
in all the fields, and $C$ corresponds to the area of a single field. 
Therefore we derive that the relative uncertainty on the number counts is given by
\begin{equation}
	\frac{\sigma_{tot}}{X_{tot}}=\sqrt{\frac{1}{X_{tot}}+\frac{AC}{N}}=0.08
\end{equation}
for the SWAN fields, while the contribution from Poisson statistics would only  
be of the order of 5\%.

We expect that the predictions of any successful model should fit the
observed SWAN counts within a discrepancy of this order of magnitude.
In good agreement with these expectations, we find that a correction
of 4\% produces the best match with the total number counts in the PLE model, 
while a correction of 1\% provides the best match to the hierarchical model. 
The model predictions discussed in the following sections have been renormalized
accordingly, i.e., PLE and hierarchical model normalizations 
are multiplied by 1.04 and 1.01 respectively.

\subsection{Comparison of the SWAN data with the model predictions}

The predictions of the PLE model for the galaxy counts and 
galaxy size as a function of the $K_\mathrm{s}$ magnitude were compared 
with the number counts and effective radii $R_e$ observed in the SWAN
images.  We are assuming for the PLE model 
$H_0=70\ \mathrm{km}\ \mathrm{s}^{-1} \mathrm{Mpc}^{-1}$,
$\Omega_m=0.27$, $\Omega_{\Lambda}=73$, and a formation redshift 
$z_F=5$ for all galaxy types (Totani et al. \cite{totani01}). 
The comparison between the PLE model
and the total completeness corrected number  
counts is shown in Fig.~\ref{tot}a. 
As explained in section~\ref{counts}, 
the points at $K_\mathrm{s} \leq 15$ show an excess due to a selection 
bias, as some fields were selected to contain more bright galaxies. 
Counts from other surveys in the literature for $K_\mathrm{s} \leq 16$ were
therefore plotted for 
comparison to the model in this range, and good agreement is found between 
the SWAN galaxy counts and the PLE predictions.

Figure~\ref{tot}b shows the average effective radius 
$\left<\log(R_e)\right>$ of the 
galaxies with the most reliable morphological fitting, i.e., those fit 
by GALFIT with $\chi^2_{\nu} \leq 2$. The average must take
into account the selection effects due to the different detection 
probabilities of the galaxies.  We therefore weighted each galaxy
using its detection probability as derived in section
\ref{counts}. The error bars show the standard error on the mean. The 
observed data points are compared with the predictions of the PLE
model, which manifests a slight overprediction of the galaxies'
$R_e$. This effect is mainly due to the late-type galaxy 
population (see Fig.~\ref{sized}), for which there might 
be hint of an increase in size, and will be discussed in section~\ref{typecomp}.
Our results confirm the finding of Totani et al. (\cite{totani01}) 
and Minowa et al. (\cite{minowa}), who found in the Subaru Deep Field and
Subaru Super Deep Field that a PLE model with no number or size
evolution gives the best fit to their $K$-selected sample's number
counts and isophotal area distribution.
\begin{figure*}
        \centering
        \begin{minipage}[c]{1.0\textwidth}
        \centerline{\hbox{
                  \psfig{file=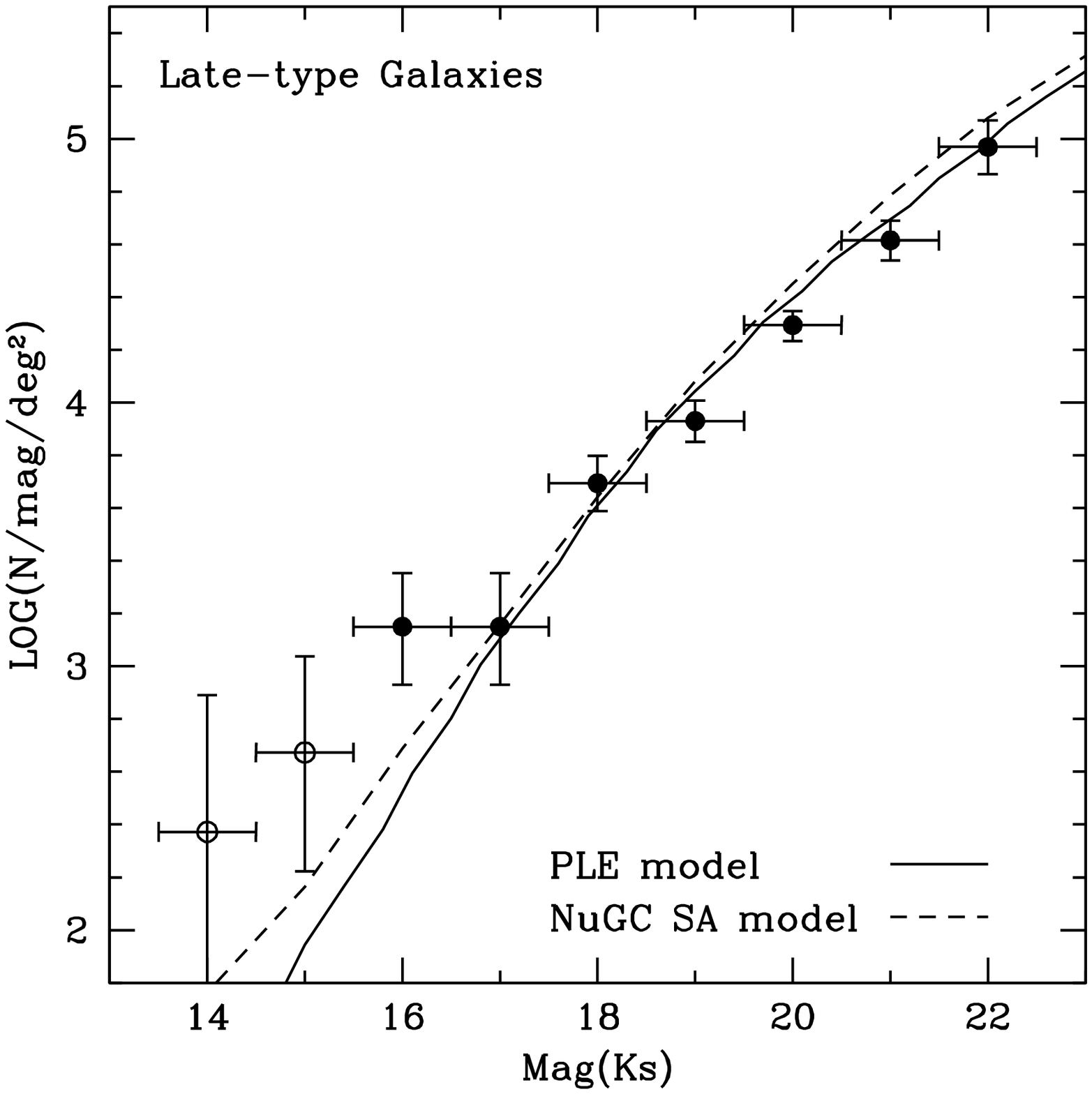,width=7.5cm}
                 \hspace{0.0cm}
                  \psfig{file=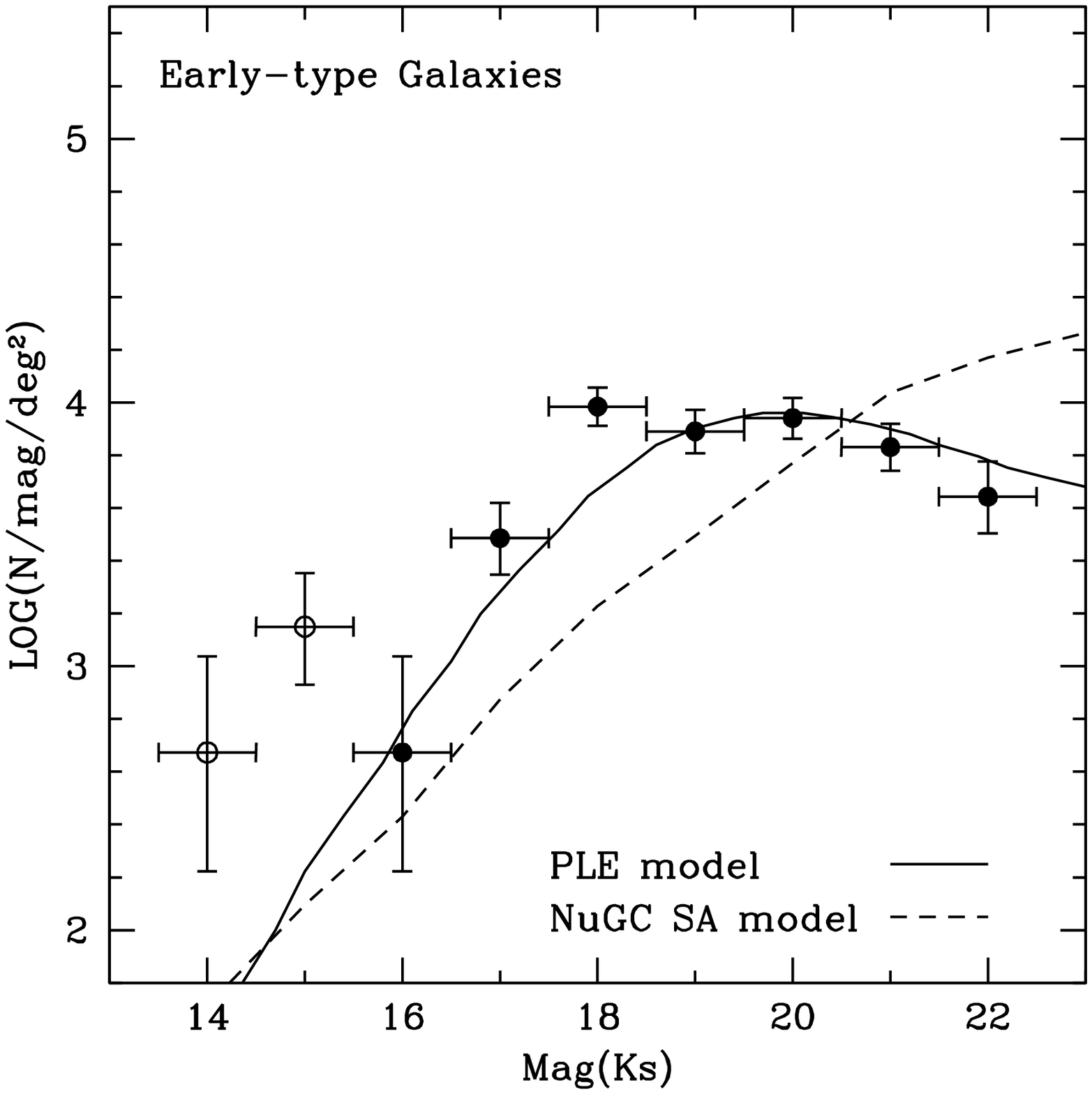,width=7.5cm}
        }}
        \caption{\label{countsd} Comparison of the SWAN 
	counts for exponential profile galaxies ($n < 2$, \textit{left panel}) 
	and early-type galaxies ($n>2$, \textit{right panel}) 
	with the PLE model (solid line) and hierarchical model (dashed line) predictions. 
	The open circles are not reliable points, as explained in the legend of 
        Fig.~\ref{tot}.}  
        \end{minipage}
\end{figure*}

The NuGC hierarchical model is also compared with the total
galaxy counts and $R_e$ distribution for SWAN in Fig.~\ref{tot}. 
Within the observational uncertainties of the shape of the observed luminosity 
functions, Nagashima et al. (\cite{naga05}) adopted two different 
models, characterized by two different supernova feedback regimes--
their strong supernova feedback (SSFB) model and weak supernova
feedback (WSFB) models.  Comparisons with the total faint number counts
and isophotal area distribution for $K$-band selected galaxies in the 
Subaru Deep Field, and with redshift distributions for faint galaxies, 
showed that the SSFB model is in better agreement with the properties
of the $K$-selected 
sample (Nagashima et al. \cite{naga02}; Nagashima et al. \cite{naga05}), 
although the predicted counts were overestimated for $K \gtrsim
23$. We therefore compare the properties of the SWAN galaxies with the SSFB 
model predictions only in the following.
A $\Lambda$CDM cosmology is used for the hierarchical model, with $\Omega_m=0.3$, 
$\Omega_{\Lambda}=0.7$ and $H_0=70\,\mathrm{\rm
  km\,s^{-1}\,Mpc^{-1}}$.  As found by 
Nagashima et al. (\cite{naga05}), the model is in good agreement with
the total number counts data, and matches the $R_e$ distribution in
SWAN as well as the PLE model.

\subsection{Morphological type dependent comparison} \label{typecomp}

In addition to the total number counts and size-magnitude relation, our  
high-resolution AO morphological classification of the SWAN galaxies 
allows us to assess the predictions of the models for the counts and
sizes of the late type and early type galaxies separately. This is 
one of the first times such a comparison has been 
done in the near-IR, as both AO
observations and accurate PSF modeling are needed to obtain 
reliable morphological classification of faint field galaxies at these 
wavelengths.  We are therefore able to compare our observational data
with untested predictions of the two models. 

In Fig. \ref{countsd}a, the PLE model predictions for the number counts of 
all galaxies with exponential profiles (spirals, irregulars and dwarf ellipticals) 
are compared with the observations
for galaxies with S\'ersic index $n<2$ according to GALFIT.  We find
good agreement between model and data for the number counts, as before.  
A very similar result is found using the hierarchical model prediction for 
late-type galaxies.

\begin{figure}
        \centering
        \psfig{file=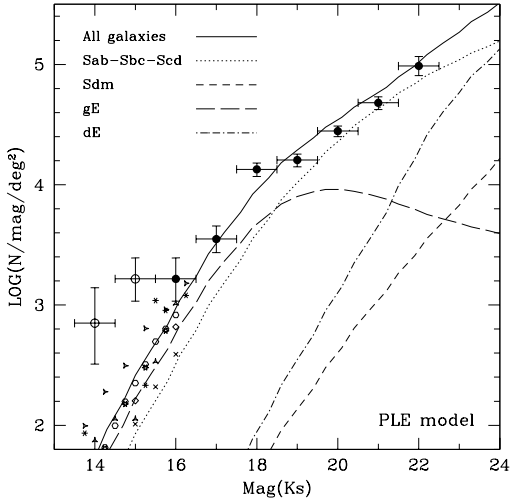,width=7.5cm}
        \caption{\label{pletutti} Comparison of the total SWAN number 
	counts with the PLE model predictions.
	The contributions of different galaxy types are shown as different 
	thin lines, while the prediction for the total population of 
	galaxies is shown as a solid line. The large open circles are not 
	reliable, as explained in the caption of Fig.~\ref{tot}.}  
\end{figure}

Fig.~\ref{countsd}b shows the predictions of both models for early-type galaxies, 
compared with the observations for $n>2$ galaxies in SWAN. In the SWAN data, 
the elliptical counts are much flatter than the late-type counts, 
showing a plateau for $K_\mathrm{s} >20$.
A much flatter slope in the early-type galaxies number counts with respect to
the late-types was found also by Teplitz et al. (\cite{teplitz}), using
HST NICMOS observations in the $H$ band, and in deep optical 
observations (e.g. Abraham et al. \cite{abraham96}). 
The adopted PLE model is able to convincingly  reproduce the plateau 
observed in the counts for $K_\mathrm{s}>20$; a similar behavior
is predicted by other PLE models (see, e.g., Pozzetti et al. 
\cite{pozzetti96}).  This plateau is produced in the model by a 
combination of two effects. First, the luminosity function of the gE
population is bell-like (see, e.g., Totani et al. \cite{totani01}),
and the number of faint gEs decreases rapidly with decreasing luminosity.
Second, in the PLE model beyond $z > 1.5$ 
giant ellipticals are very faint, due to heavy extinction in the model 
($\tau \gtrsim 10$), as described in Totani \& Yoshii (\cite{totani00}). 
As a result, for $K_s \geq 20$ (corresponding to an $L^*$ galaxy at $z=1.5$), 
going to fainter magnitudes does not correspond to an increase of the sampled 
volume. Instead, beyond $K_\mathrm{s} = 20$, the model predicts we
should begin to miss the dusty high-redshift progenitors of 
today's ellipticals, consistent with the plateau observed in the SWAN data. 
A scenario in which massive $z \gtrsim 1.5$ ellipticals are highly 
obscured by dust during their starburst phase, and therefore produce 
the plateau observed in our $K$-band number counts, is consistent with 
the detection of very luminous, highly obscured submillimeter galaxies
at high redshift (Blain et al. \cite{blain02}, and references
therein). In addition, galaxies with unusually red IR colors that have been 
measured in deep 
in near-IR observations can be explained as primordial elliptical 
galaxies that are 
reddened by dust and still in the starburst phase of their formation 
at $z \gtrsim 3$ (e.g., Totani et al. \cite{totani01b}).

In contrast to the success of the PLE model, the hierarchical considerably underpredicts 
at bright magnitudes ($K_\mathrm{s} \lesssim 20$) and overpredicts 
at fainter magnitudes the observed number counts.  In particular, the observed plateau in the 
early-type counts, which was very well reproduced in the PLE model,  
is not expected at all in the hierarchical model predictions. 
This disagreement implies that the processes 
that produce an elliptical galaxy, in at least this particular 
hierarchical model, are not adequate to describe reality. 
In the model, an elliptical galaxy 
is formed through a major merger, i.e., 
a merger with mass ratio $f=m_1/m_2 \geq 0.3$, in which  
it is assumed that all the cold gas turns into stars and 
hot gas, and all the stars populate the bulge of a new galaxy.
Although it may be possible to increase the number of bright ellipticals by 
changing the model parameter that regulates the mass ratio
distinguishing major from minor mergers, 
it seems harder to decrease the model's number of $K_\mathrm{s} > 20$ 
ellipticals to the level observed in SWAN  
(Nagashima private communication). In this case no separation is 
possible between the gE and dE populations, 
as all the major mergers produce galaxies with a 
de Vaucouleurs profile. 

We note that misclassification of galaxies between the two 
categories of early and late-type is not expected to strongly affect 
these conclusions. We expect only $\sim 10\%$ of late-type to be 
misclassified as early-type and $\sim 10\%$ of early-type 
to be misclassified as late-type at $K_\mathrm{s} = 21$ (see Paper I). Since
at faint magnitudes the number counts of spirals are $\sim 10$ 
times higher than those of ellipticals, it is more likely that 
we are overestimating the number of faint ellipticals in our sample. 
Correcting for this bias would make the discrepancy with 
the hierarchical model even larger.

\begin{figure*}
        \centering
        \begin{minipage}[c]{1.0\textwidth}
        \centerline{\hbox{
                  \psfig{file=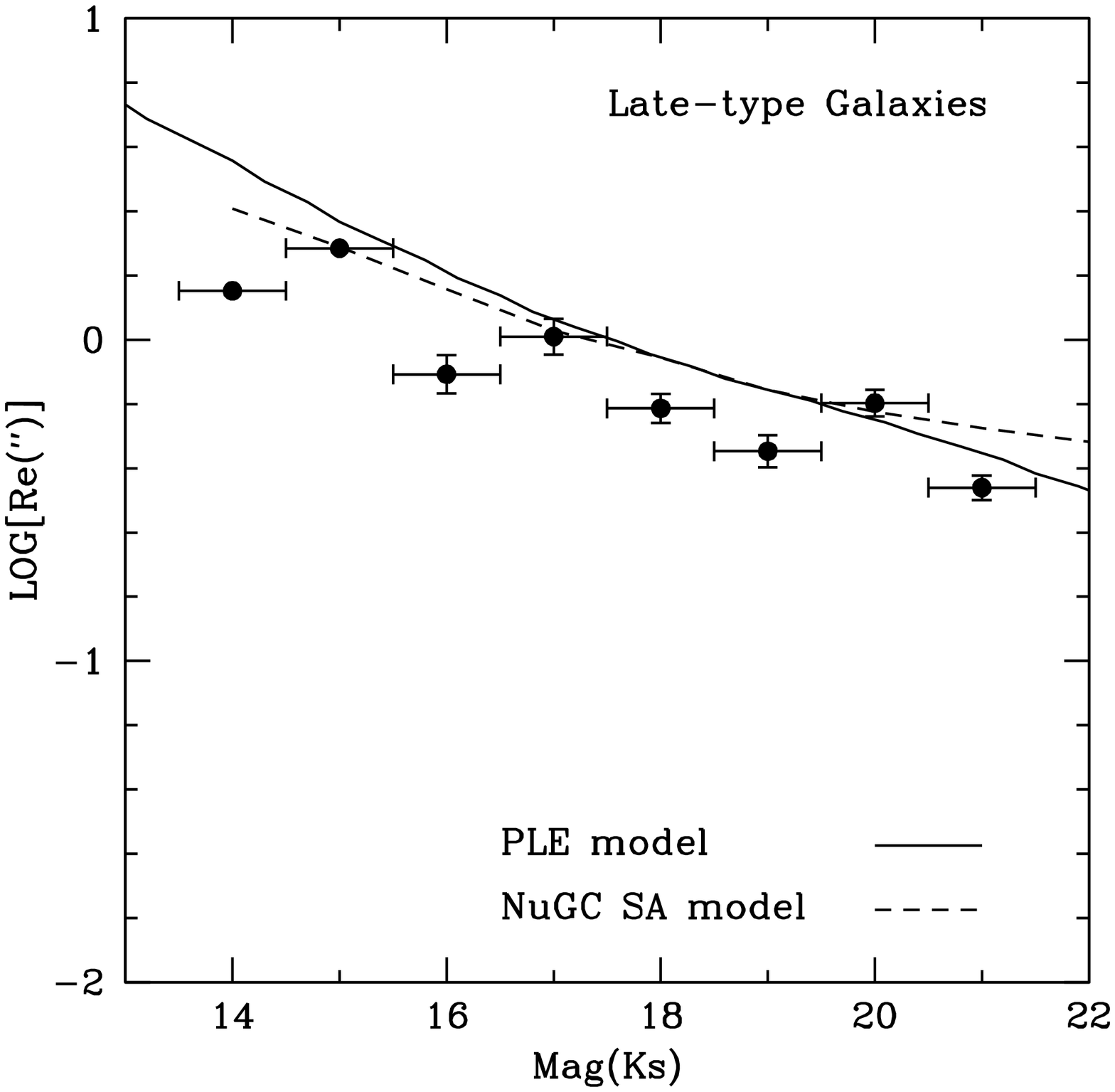,width=7.5cm}
                 \hspace{0.0cm}
                  \psfig{file=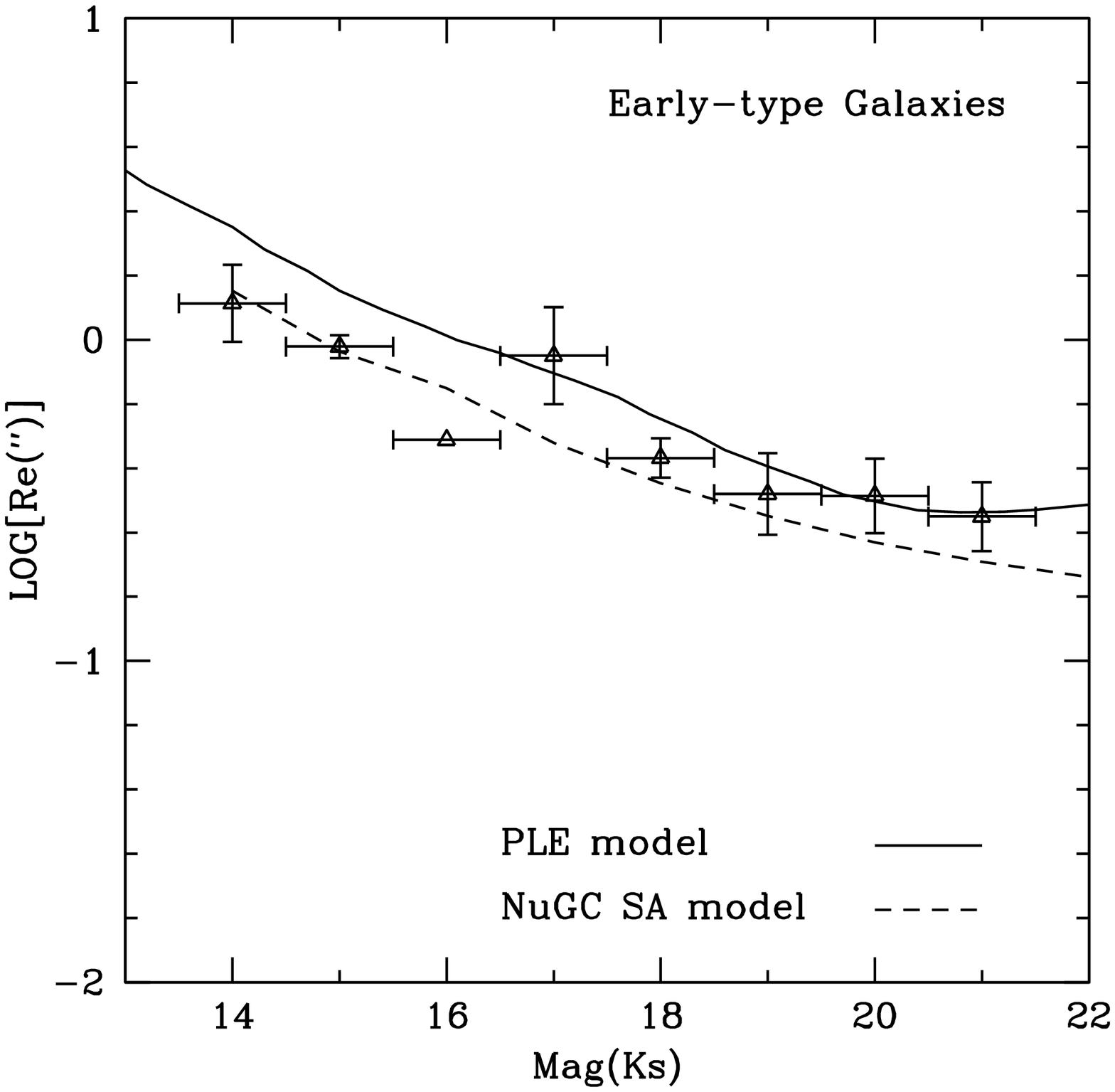,width=7.5cm}
        }}
        \caption{\label{sized} Comparison of the  
	$R_e$ distribution as a function of magnitude of late-type (\textit{left panel}) 
	and early-type galaxies (\textit{right panel})  
	with $\chi^2_{\nu} \leq2$ in the GALFIT fit
	with the PLE model (solid line) and hierarchical model (dashed line) 
        predictions.  The error bars are the standard error on the mean; no 
	error bar is drawn for a bin with one galaxy.}  
        \end{minipage}
\end{figure*}

We conclude that the PLE model better reproduces the observed number 
counts of the SWAN galaxies. In Fig.~\ref{pletutti} the contributions 
of the different galaxy types to the total observed number counts 
according to the PLE model are shown. The separation between the two 
populations of dE and gE galaxies  
proves to be a key element in reproducing the number counts separated by 
morphological type on the basis of their light profiles. In fact, 
the number counts 
for exponential-like profiles would have been underestimated for 
$K_\mathrm{s} \gtrsim 20$, and the $r^{1/4}$ profiles number counts 
overestimated, had the dE 
galaxies been included 
in the early-type morphological bin along with the gE galaxies. 
This result confirms 
the suggestions 
of Totani et al. (\cite{totani01}) that such a separation better reproduced 
the observed number counts in the Subaru Deep Field for $K\gtrsim20$ than 
a model with a single elliptical population. \\

In Fig.~\ref{sized}, the observed size-magnitude distribution discriminating between late-type 
and early-type galaxies is compared with the model predictions.  At bright magnitudes, 
the observed galaxies are smaller than the model predictions for both early-type and 
late-type samples.  This discrepancy is likely due to an uncorrected systematic effect: 
large ($R_e \gtrsim 1\arcsec$) bright galaxies can be self-subtracted in the reduction process, as 
explained in section~\ref{reduc}.  This phenomenon will reduce the apparent sizes of sources, 
although their integrated magnitudes (and thus their contributions to the number 
counts) will be only minimally affected.

At faint magnitudes, the models are able to reproduce the observed distributions for 
early-type galaxies. However, the PLE model better reproduces the observed distribution 
at fainter magnitudes, where the hierarchical model predicts more compact objects
than are observed. In contrast, an offset may exist 
between both models and the data for late-type galaxies, which appear $\sim 30\%$ smaller 
than predicted at faint magnitudes.  This offset is what would be expected 
for modest growth in the sizes of late-type galaxies.  In both of the models considered 
here, the sizes of disks for a given mass are almost independent of the formation redshift.  
This property is built into any PLE model, but even the hierarchical model by Nagashima 
et al. (\cite{naga05}) assumes that there is almost no evolution in the stellar mass-size relation 
for disk galaxies, as suggested by the observations of Barden et al. (\cite{barden05}) up to $z \sim 1$.  
If our observed offset is really due to increses in size in the late-type population, the result 
would be qualitatively consistent e.g. with the predictions of Mo et al. (\cite{mo98}), who estimated 
that disk galaxies forming at $z = 1$ are 50\% smaller than disks forming at $z = 0$.  
However, the inclusion of galaxies with many different redshifts, masses, 
and $M/L$ ratios in the faint bins of Fig.~\ref{sized}b prevents any robust quantitative 
conclusion. \\

Our finding that pure luminosity evolution of galaxies is favored for a 
$K_\mathrm{s}$-selected sample up to $K_\mathrm{s} \sim 22$, 
without evidence of relevant number evolution even when separating between late 
and early-type galaxies is consistent with other results. For example,   
Truijllo et al. (\cite{trujillo05}) used deep near-infrared images 
from the FIRES survey, combined with GEMS and SDSS data, to confirm that 
the observed size-magnitude relation evolution out to $z\sim1.7$ for late-type objects
matches very well the expected evolution for Milky-Way type
objects from infall models, while 
for spheroid-like objects the evolution of the
luminosity-size relation was found to be consistent with pure luminosity 
evolution of a fading galaxy population.  
McIntosh et al. (\cite{mcintosh05}) studied a large sample of early-type 
galaxies from the GEMS survey, finding that the luminosity-size and stellar mass-size 
relations evolve in a manner that is consistent with the passive aging of ancient  
stellar population.
Papovich et al. (\cite{papovich05}) suggest that passive evolution 
can account for the observed luminosity-size relation
at $z\sim 1$, with merging becoming important at higher redshifts.

\section{Conclusions} \label{concl}

In this paper we have presented new results from a high resolution 
adaptive optics assisted morphological study of 21 fields from SWAN,
the Survey of a Wide Area with NACO.  The PSF model derived in Paper I 
was used in combination with GALFIT to classify the SWAN galaxies into 
the two classes of early and late type, and to derive effective radii 
$R_e$ of 383 galaxies. A detailed study of the detection probability
as a function of the magnitude, S\'ersic index, effective radius,
field and distance from the guide star was performed in order 
to take careful account of the selection effects affecting our sample. 
The results were used to compute the completeness-corrected number counts 
and to derive the average $R_e$ as a function of magnitude.

The number counts and size-magnitude relation for the total galaxy population,  
and for early and late-type separately, were compared with both 
a modified version of the pure luminosity evolution model of 
Totani \& Yoshii (\cite{totani00}) and with the {\it a priori}
hierarchical model developed by Nagashima et al. (\cite{naga05}). 
We have shown in section~\ref{compare} that while the hierarchical model 
can convincingly reproduce the counts of late-type galaxies, 
it is not consistent with the observed number counts of elliptical galaxies 
selected in the $K_\mathrm{s}$ band. On the other hand, the PLE model 
can reasonably reproduce both the late and early type count  
distributions for the SWAN galaxies. 
We have compared the size-magnitude distribution of the galaxies 
with the predictions of the models, finding that there might be some hint 
of increased size for the late-type galaxy population. Both models are consistent 
with the observed distribution for early-type galaxies, although the PLE model 
seems to better reproduce the observed distribution at fainter magnitudes.
Our work therefore favors pure luminosity evolution of early-type galaxies for a 
$K_\mathrm{s}$-selected sample up to $K_\mathrm{s} \sim 22$. In contrast, our results show 
that a representative example of currently available models based on the hierarchical galaxy 
formation theory is not able to reproduce the observed properties of faint
$K_\mathrm{s}$-selected early-type galaxies in the near-IR. 

These results illustrate the importance of obtaining reliable
morphological classifications for better constraining the details of
galaxy formation and evolution models, and demonstrate the unique
power of AO observations to extend such work to faint galaxies in the near-IR.
  

\begin{acknowledgements}

We thank the anonymous referee for useful comments and suggestions.
The authors are grateful to the staff at Paranal Observatory for their
hospitality and support during the observations. We thank 
Masahiro Nagashima for 
providing the $K$-band simulated data from the Numerical 
Galaxy Catalog and useful discussion of our results;  
Reinhard Genzel, Reiner Hofmann, Sebastian Rabien, 
Niranjan Thatte, and W. Jimmy Viehhauser, for their help and 
discussion of SWAN strategy and results; and Rainer 
Sch\"odel for the observations of SBSF\,41.
Some of the data included in this paper were obtained as part of the MPE
guaranteed time programme. GC and AJB acknowledge MPE for support; 
AJB acknowledges support from 
the National Radio Astronomy Observatory, which is operated by Associated 
Universities, Inc., under cooperative agreement with the National 
Science Foundation.
\end{acknowledgements}


\end{document}